\documentclass[final]{cvpr}

\usepackage{times}
\usepackage{epsfig}
\usepackage{graphicx}
\usepackage{amsmath}
\usepackage{amssymb}
\usepackage{booktabs}
\usepackage{multirow}
\usepackage{url}

\usepackage[pagebackref=true,breaklinks=true,colorlinks,bookmarks=false]{hyperref}

\def\cvprPaperID{****} 
\def\confYear{CVPR 2023}

\begin{document}

\title{MIPI 2023 Challenge on RGBW Fusion: Methods and Results}

\author{
Qianhui Sun \and Qingyu Yang \and Chongyi Li \and Shangchen Zhou \and Ruicheng Feng \and Yuekun Dai \and Wenxiu Sun \and Qingpeng Zhu \and Chen Change Loy \and Jinwei Gu \and
Hongyuan Yu \and Yuqing Liu \and Weichen Yu \and Lin Ge \and Xiaolin Zhang \and Qi Jia \and Heng Zhang \and Xuanwu Yin \and Kunlong Zuo \and
Qi Wu \and Wenjie Lin \and Ting Jiang \and Chengzhi Jiang \and Mingyan Han \and Xinpeng Li \and Jinting Luo \and Lei Yu \and Haoqiang Fan \and Shuaicheng Liu \and
Kunyu Wang \and Chengzhi Cao \and
Yuanshen Guan \and Jiyuan Xia \and Ruikang Xu \and Mingde Yao \and Zhiwei Xiong
}

\maketitle

\begin{abstract}
Developing and integrating advanced image sensors with novel algorithms in camera systems are prevalent with the increasing demand for computational photography and imaging on mobile platforms. 
However, the lack of high-quality data for research and the rare opportunity for an in-depth exchange of views from industry and academia constrain the development of mobile intelligent photography and imaging (MIPI). 
With the success of the \href{https://mipi-challenge.org/MIPI2022/}{1st MIPI Workshop@ECCV 2022}, we introduce the second MIPI challenge, including four tracks focusing on novel image sensors and imaging algorithms.
This paper summarizes and reviews the RGBW Joint Fusion and Denoise track on MIPI 2023.
In total, 69 participants were successfully registered, and 4 teams submitted results in the final testing phase.
The final results are evaluated using objective metrics, including PSNR, SSIM, LPIPS, and KLD. 
A detailed description of the models developed in this challenge is provided in this paper. 
More details of this challenge and the link to the dataset can be found at \href{https://mipi-challenge.org/MIPI2023/}{https://mipi-challenge.org/MIPI2023/}.
\end{abstract}

{\let\thefootnote\relax\footnotetext{%
\tiny  Qianhui Sun$^{1}$ (\href{sunqianhui@sensebrain.site}{sunqianhui@sensebrain.site}), Qingyu Yang$^{1}$ (\href{yangqingyu@sensebrain.site}{yangqingyu@sensebrain.site}), Chongyi Li$^{4}$, Shangchen Zhou$^{4}$, Ruicheng Feng$^{4}$, Wenxiu Sun$^{2,3}$, Qingpeng Zhu$^{2}$, Chen Change Loy$^{4}$, Jinwei Gu$^{1,3}$ are the MIPI 2023 challenge organizers
($^{1}$SenseBrain, $^{2}$SenseTime Research and Tetras.AI, $^{3}$Shanghai AI Laboratory, $^{4}$Nanyang Technological University). The other authors participated in the challenge. Please refer to Appendix~\ref{appendix:teams} for details.
\\
MIPI 2023 challenge website: \href{https://mipi-challenge.org/MIPI2023/}{https://mipi-challenge.org/MIPI2023/}
}
}


\section{Introduction}
\begin{figure}[!ht]
\centering
\includegraphics[width=0.5\textwidth]{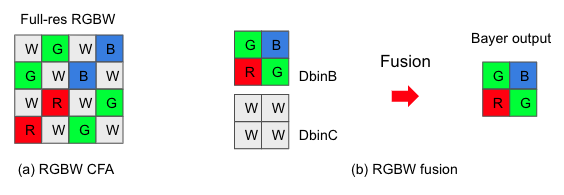}
\caption{The RGBW Fusion task: (a) the RGBW CFA. (b) In the binning mode, DbinB and DbinC are obtained by diagonal averaging of pixels of the same color within a 2$\times$2 window. The joint fusion and denoise algorithm takes DbinB and DbinC as input to get a high-quality Bayer.}
\label{fig:rgbw_cfa}
\setlength{\belowcaptionskip}{0pt plus 3pt minus 2pt}
\end{figure}

RGBW is a new type of CFA (Color Filter Array) pattern (Fig.~\ref{fig:rgbw_cfa} (a)) designed for image quality enhancement under low light conditions. Thanks to the higher optical transmittance of white pixels over conventional red, green, and blue pixels, the signal-to-noise ratio (SNR) of images captured by this type of sensor increases significantly, thus boosting the image quality, especially under low light conditions. Recently, several phone OEMs~\cite{TranssionRGBW, oppoRGBW, vivoRGBW} have adopted RGBW sensors in their flagship smartphones to improve the camera image quality.

The binning mode of RGBW is mainly used in the camera preview mode and video mode, in which a half-resolution Bayer is generated from the RGBW image, where spatial resolution is traded off for faster response. In this mode, every two pixels of the same color within a $2\times2$ window of the RGBW are averaged in the diagonal direction, and a diagonal-binning-Bayer image (DbinB) and a diagonal-binning-white image (DbinC) are generated. A fusion algorithm is demanded to enhance details and reduce noise in the Bayer image with the help of the white image (Fig.~\ref{fig:rgbw_cfa} (b)). A good fusion algorithm should be able to fully take advantage of the SNR and resolution benefit of white pixels.

The RGBW fusion problem becomes more challenging when the input DbinB and DbinC become noisy, especially under low light conditions. A joint fusion and denoise task is thus in demand for real-world applications.

\begin{figure*}[!ht]
\centering
\includegraphics[width=0.9\textwidth]{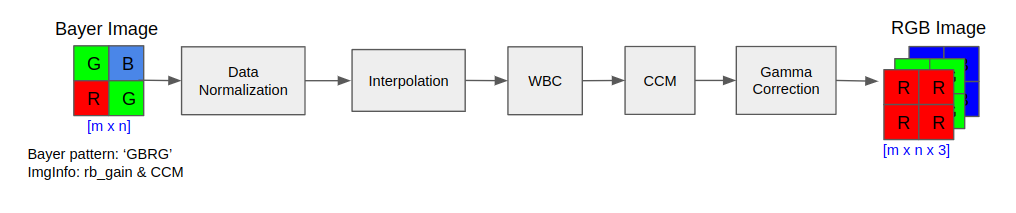}
\caption{An ISP to visualize the output Bayer and to calculate the loss function.}
\label{fig:simple_isp}
\setlength{\belowcaptionskip}{0pt plus 3pt minus 2pt}
\end{figure*}

In this challenge, we intend to fuse DbinB and DbinC in Fig.~\ref{fig:rgbw_cfa} (b) to denoise and improve the Bayer. The solution is not necessarily learning-based. However, we provide a high-quality dataset of binning-mode RGBW input (DbinB and DbinC) and the output Bayer pairs to facilitate learning-based methods development, including 100 scenes (70 scenes for training, 15 for validation, and 15 for testing). The dataset is similar to the one provided in the first MIPI challenge, while we replaced some similar scenes with new ones. We also provide a simple ISP for participants to get the RGB image results from Bayer for quality assessment. Fig.~\ref{fig:simple_isp} shows the pipeline of the simple ISP. The participants are also allowed to use other public-domain datasets. The algorithm performance is evaluated and ranked using objective metrics: Peak Signal-to-Noise Ratio (PSNR), Structural Similarity Index (SSIM)~\cite{ssim}, Learned Perceptual Image Patch Similarity (LPIPS)~\cite{lpips}, and KL-divergence (KLD). 

We hold this challenge in conjunction with the second MIPI Challenge which will be held on CVPR 2023. Similar to the first MIPI challenges~\cite{feng2023mipi,sun2023mipi,yang2023mipi,yang2023mipi2,yang2023mipi3}, we are seeking algorithms that fully take advantage of the SNR and resolution benefit of white pixels to enhance the final Bayer image in the binning model. MIPI 2023 consists of four competition tracks:

\begin{itemize}
    \item \textbf{RGB+ToF Depth Completion} uses sparse and noisy ToF depth measurements with RGB images to obtain a complete depth map.
    \item \textbf{RGBW Sensor Fusion} fuses Bayer data and a monochrome channel data into Bayer format to increase SNR and spatial resolution.
    \item \textbf{RGBW Sensor Remosaic} converts RGBW RAW data into Bayer format so that it can be processed by standard ISPs.
    \item \textbf{Nighttime Flare Removal} is to improve nighttime image
quality by removing lens flare effects.
\end{itemize}


\section{MIPI 2023 RGBW Sensor Fusion}

To facilitate the development of high-quality RGBW fusion solutions, we provide the following resources for participants:
\begin{itemize}
    \item A high-quality dataset of aligned RGBW (DbinB and DbinC in Fig.~\ref{fig:rgbw_cfa} (b)) and Bayer. We enriched the scenes compared to the first MIPI challenge dataset. As far as we know, this is the only dataset consisting of aligned RGBW and Bayer pairs;
    \item A script that reads the provided raw data to help participants get familiar with the dataset;
    \item A simple ISP including basic ISP blocks to visualize the algorithm outputs and to evaluate image quality on RGB results;
    \item A set of objective image quality metrics to measure the performance of a developed solution.
\end{itemize}

\subsection{Problem Definition}
The RGBW fusion task aims to fuse the DbinB and DbinC of RGBW (Fig.~\ref{fig:rgbw_cfa} (b)) to improve the image quality of the Bayer output. By incorporating the white pixels in DbinC of higher spatial resolution and higher SNR, the output Bayer would potentially have better image quality. In addition, the binning mode of RGBW is mainly used for the preview and video modes in smartphones, thus requiring the fusion algorithms to be lightweight and power-efficient. While we do not rank solutions based on the running time or memory footprint, the computational cost is one of the most important criteria in real applications.

\subsection{Dataset: Tetras-RGBW-Fusion}
\begin{figure}[!ht]
\centering
\includegraphics[width=0.5\textwidth]{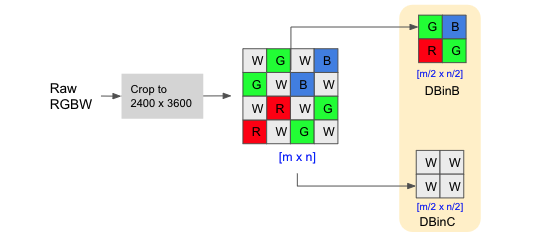}
\caption{Data generation of the RGBW fusion task. The RGBW raw data is captured using an RGBW sensor and cropped into a size of $2400\times3600$. A Bayer (DbinB) and white (DbinC) image are obtained by averaging the same color in the diagonal direction within a $2\times2$ block.}
\label{fig:data_gen}
\setlength{\belowcaptionskip}{0pt plus 3pt minus 2pt}
\end{figure}

The training data contains 70 scenes of aligned RGBW (DbinB and DbinC input) and Bayer (ground-truth) pairs. DbinB at 0dB is used as the ground truth for each scene. Noise is synthesized on the 0dB DbinB and DbinC data to provide the noisy input at 24dB and 42dB, respectively. The synthesized noise consists of read noise and shot noise, and the noise models are calibrated on an RGBW sensor. The data generation steps are shown in Fig.~\ref{fig:data_gen}. The testing data includes DbinB and DbinC inputs of 15 scenes at 24dB and 42dB, and the ground truth Bayer results are hidden from participants during the testing phase.

\subsection{Evaluation}
The evaluation consists of (1) the comparison of the fusion output Bayer and the reference ground truth Bayer, and (2) the comparison of RGB from the predicted and ground truth Bayer using a simple ISP (the code of the simple ISP is provided). We use
\begin{enumerate}
    \item Peak Signal-to-Noise Ratio (PSNR)
    \item Structural Similarity Index Measure (SSIM)~\cite{ssim}
    \item Learned Perceptual Image Patch Similarity (LPIPS)~\cite{lpips}
    \item Kullback–Leibler Divergence (KLD)
\end{enumerate}
to evaluate the fusion performance. The PSNR, SSIM, and LPIPS will be applied to the RGB from the Bayer using the provided simple ISP code, while KLD is evaluated on the predicted Bayer directly.

A metric weighting PSNR, SSIM, KLD, and LPIPS is used to give the final ranking of each method, and we will report each metric separately as well. The code to calculate the metrics is provided. The weighted metric is shown below. The M4 score is between 0 and 100, and the higher score indicates the better overall image quality.

\begin{equation}
    M4 = PSNR \cdot SSIM \cdot 2^{1-LPIPS-KLD}.
\label{eq:M4}
\end{equation}
For each dataset, we report the average score over all the processed images belonging to it.

\subsection{Challenge Phase}
The challenge consisted of the following phases:
\begin{enumerate}
    \item Development: The registered participants get access to the data and baseline code, and are able to train the models and evaluate their running time locally.
    \item Validation: The participants can upload their models to the remote server to check the fidelity scores on the validation dataset, and to compare their results on the validation leaderboard.
    \item Testing: The participants submit their final results, code, models, and factsheets.
\end{enumerate}


\section{Challenge Results}

Four teams submitted their results in the final phase, which have been verified using their submitted code. Table.~\ref{tab:results} summarizes the results in the final test phase. \textbf{RUSH MI}, \textbf{MegNR}, and \textbf{USTC-Zhalab} are the top three teams ranked by M4 are presented in Eq.~\eqref{eq:M4}, and \textbf{RUSH MI} shows the best overall performance. The proposed methods are described in Section \ref{sec:methods}, and the team members and affiliations are listed in Appendix \ref{appendix:teams}.

\begin{table}[!ht]  
    \centering
    \scalebox{0.9}{
        \begin{tabular}{l | llll | l}
            \hline
            \textbf{Team name} & \textbf{PSNR} & \textbf{SSIM} & \textbf{LPIPS} & \textbf{KLD} & \textbf{M4}          \\ 
            \hline  \hline
            \text{RUSH MI}     & 38.587        & 0.977         & 0.0661         & 0.0718       & \textbf{68.58}       \\ 
            \hline
            \text{MegNR}       & 37.822        & 0.966         & 0.0815         & 0.0717       & \textbf{65.84}       \\ 
            \hline
            \text{USTC-Zhalab} & 37.323        & 0.965         & 0.0854         & 0.0767       & \textbf{64.67}       \\ 
            \hline
            \text{VIDAR}       & 37.160        & 0.968         & 0.1023         & 0.0698       & 63.98                \\ 
            \hline
        \end{tabular}
    }
    \caption{MIPI 2023 Joint RGBW Fusion and Denoise challenge results and final rankings. PSNR, SSIM, LPIPS, and KLD are calculated between the submitted results from each team and the ground truth data. A weighted metric, M4, by Eq.~\eqref{eq:M4} is used to rank the algorithm performance, and the top three teams with the highest M4 are highlighted.  
    \label{tab:results}}
\end{table}

\begin{figure*}[!ht]
\setlength{\abovecaptionskip}{0.cm}
\setlength{\belowcaptionskip}{-0.cm}
\centering
\includegraphics[width=\textwidth]{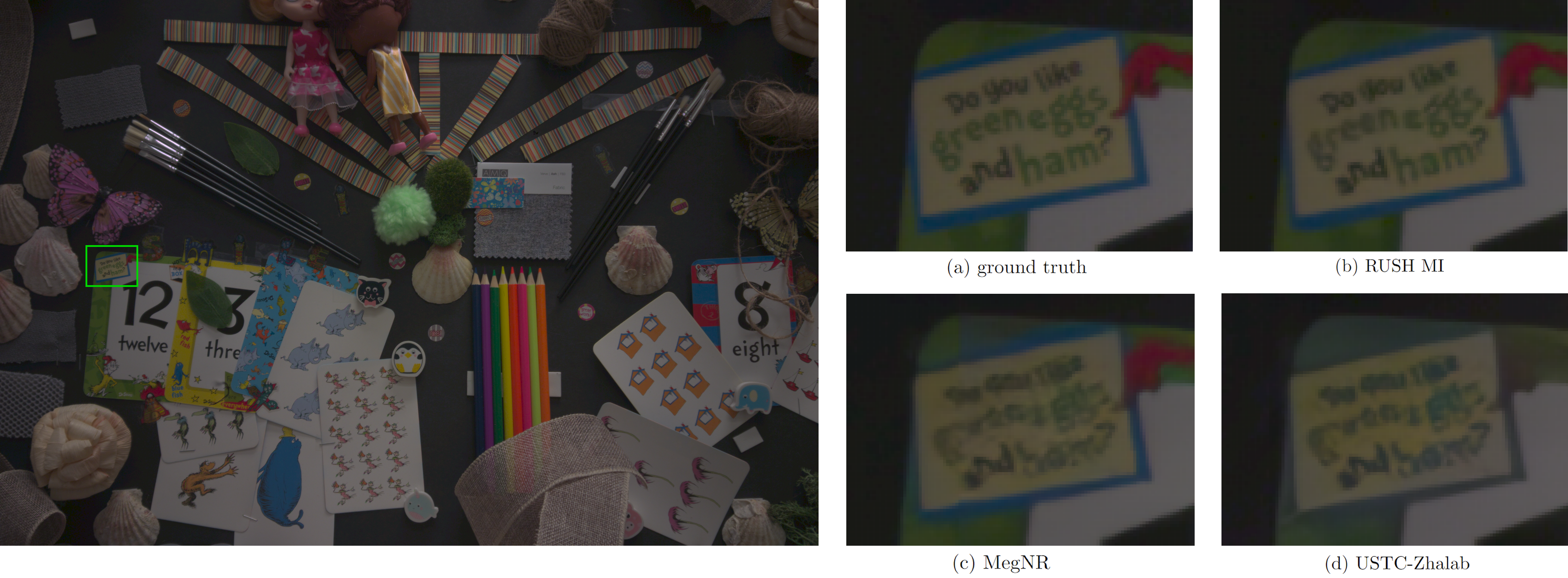}
\caption{Qualitative image quality (IQ) comparison. The results of one of the test scenes (42dB) are shown. While the top three fusion methods achieve high objective IQ metrics in Table~\ref{tab:results}, texts on the card are slightly blurred in (b) and are barely interpretable in (c) and (d). The RGB images are obtained by using the ISP in Fig.~\ref{fig:simple_isp}, and its code is provided to participants.}
\label{fig:IQ1}
\end{figure*}

\begin{figure*}[!ht]
\setlength{\abovecaptionskip}{0.cm}
\setlength{\belowcaptionskip}{-0.cm}
\centering
\includegraphics[width=\textwidth]{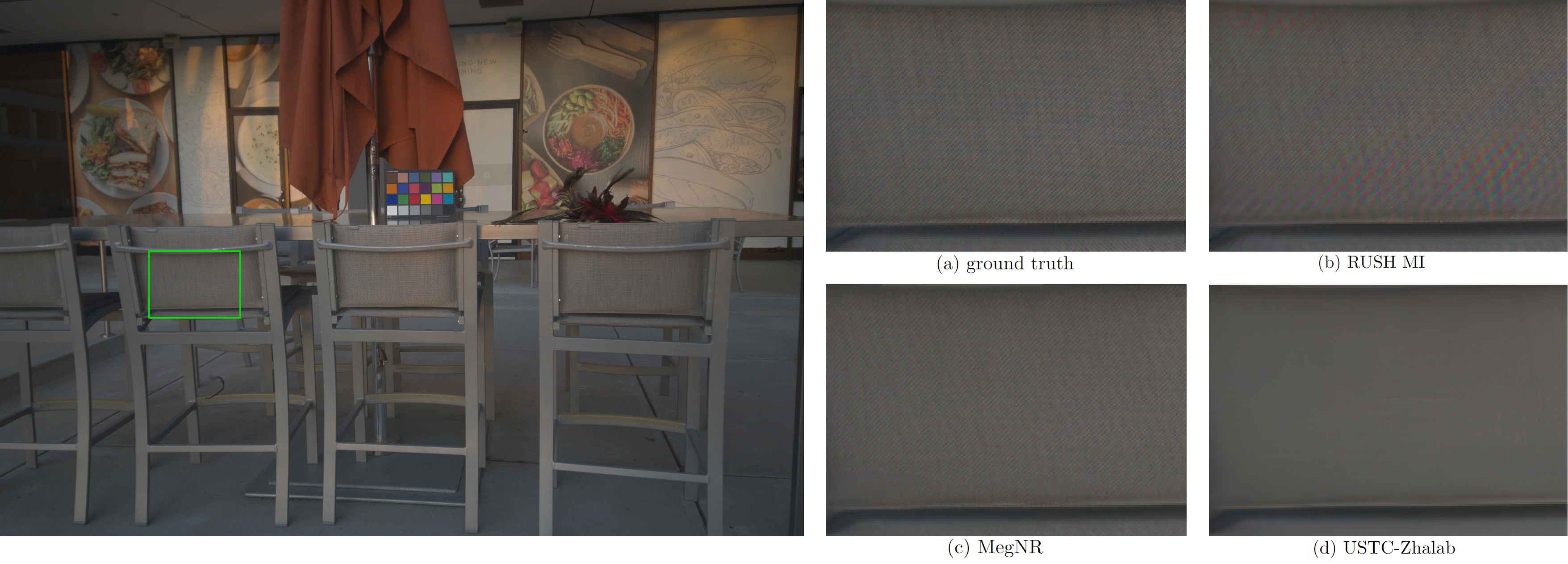}
\caption{Qualitative image quality (IQ) comparison. The results of one of the test scenes (42dB) are shown. Detail loss or false color in the top three methods in Table~\ref{tab:results} can be found when compared with the ground truth. The mesh of the chair is over-smoothed to different extents in (c) and (d) and some false color can be found in (b). The RGB images are obtained by using the ISP in Fig.~\ref{fig:simple_isp}, and its code is provided to participants.}
\label{fig:IQ2}
\end{figure*}

To learn more about the algorithm performance, we evaluated the qualitative image quality in Fig.~\ref{fig:IQ1} and Fig.~\ref{fig:IQ2} in addition to the objective IQ metrics. While all teams in Table~\ref{tab:results} have achieved high PSNR and SSIM, detail loss can be found on the texts of the card in Fig.~\ref{fig:IQ1} and detail loss or false color can be found on the mesh of the chair in Fig.~\ref{fig:IQ2}. When the input has a large amount of noise, oversmoothing tends to yield higher PSNR at the cost of detail loss perceptually.

\begin{table}[!ht]  
    \centering
    \scalebox{0.9}{
        \begin{tabular}{l | l | l}
        \hline
            \textbf{Team name} & \textbf{1200$\times$1800 (measured)} & \textbf{16M} (estimated) \\
            \hline  \hline
            \text{RUSH MI}     & \textbf{0.45s}                       &  \textbf{3.33s} \\
            \hline
            \text{MegNR}       & 8.46s                                &  62.60s \\
            \hline
            \text{USTC-Zhalab} &  69.60s                               &  515.04s \\
            \hline
        \end{tabular}
    }
    \caption{Running time of the top three solutions ranked by Eq.~\eqref{eq:M4} in the MIPI 2023 Joint RGBW Fusion and Denoise challenge. The running time of input of $1200\times1800$ was measured, while the running time of a 64M RGBW sensor was based on estimation (the binning-mode resolution of a 64M RGBW sensor is 16M).  The measurement was taken on an NVIDIA Tesla V100-SXM2-32GB GPU.
    \label{tab:runtime}}
\end{table}

In addition to benchmarking the image quality of fusion algorithms, computational efficiency is evaluated because of the wide adoption of RGBW sensors in smartphones. We measured the running time of the RGBW fusion solutions of the top three teams in Table~\ref{tab:runtime}. While running time is not employed in the challenge to rank fusion algorithms, the computational cost is critical when developing smartphone algorithms. RUSH MI achieved the shortest running time among the top three solutions on a workstation GPU (NVIDIA Tesla V100-SXM2-32GB). With the sensor resolution of mainstream smartphones reaching 64M or even higher, power-efficient fusion algorithms are highly desirable.


\section{Challenge Methods}\label{sec:methods}

This section describes the solutions submitted by all teams participating in the final stage of the MIPI 2023 RGBW Joint Fusion and Denoise Challenge. 

\subsection{RUSH MI}

\begin{figure}[!ht]
    \centering
    \includegraphics[width=\linewidth]{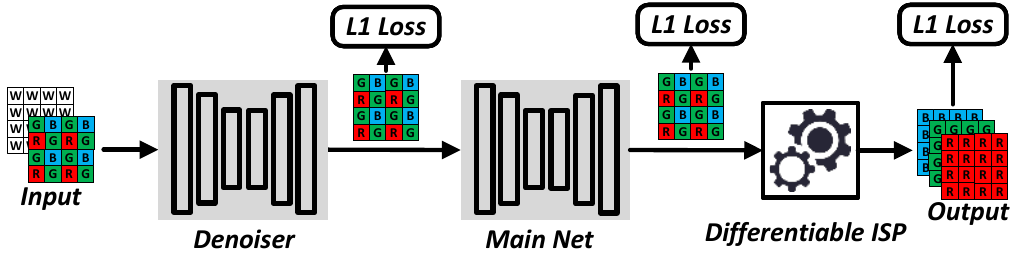}
    \caption{The model architecture of RUSH MI.}
    \label{fig:DDED}
\end{figure}

This team presents an end-to-end joint remosaic and denoise model, referred to as DEDD. As illustrated in Fig.~\ref{fig:DDED}, the DEDD model is composed of three components: a denoising model, a main network, and a differentiable ISP model. Specifically, in the first part, they employed a basic UNet~\cite{ronneberger2015u}, which incorporated two downsampling operations. In the second part, they utilized the state-of-the-art model in the low-level domain, NAFNet~\cite{chen2022simple}. The NAFNet contains the 4-level encoder-decoder and bottleneck. For the encoder, the numbers of NAFNet’s blocks for each level are 2, 4, 8, and 24. For the decoder, the numbers of NAFNet’s blocks for the 4 levels are all 2. In addition, the number of NAFNet’s blocks for the bottleneck is 12.
In the third part, they reformulated the BLC, WBC, GAMMA, and CCM modules in the conventional ISP pipeline into differentiable models, and adopted the officially provided demosaic model as the demosaic module. In the training phase, the clean model is used as a guidance for boosting the
noisy restoration performance. The images were randomly cut into 256$\times$256 patches, with a batch size of 64. The optimizer used was AdamW~\cite{loshchilov2017decoupled}, and the initial learning rate was set to 0.001, which was reduced by half every 5000 iterations. The training process is divided into two stages; initially, the denoising network is trained for 40k iterations, after which the parameters of the denoiser are fixed and the demosaic network is trained. The loss function utilized is the L1 loss. When training the demosaic network, two supervision signals are incorporated: the constraint of the Bayer domain and the constraint of the RGB domain. The model training is completed after retraining 80K iterations in an end-to-end training mode.

\subsection{MegNR}

\begin{figure}[!ht]
    \centering
    \includegraphics[width=\linewidth]{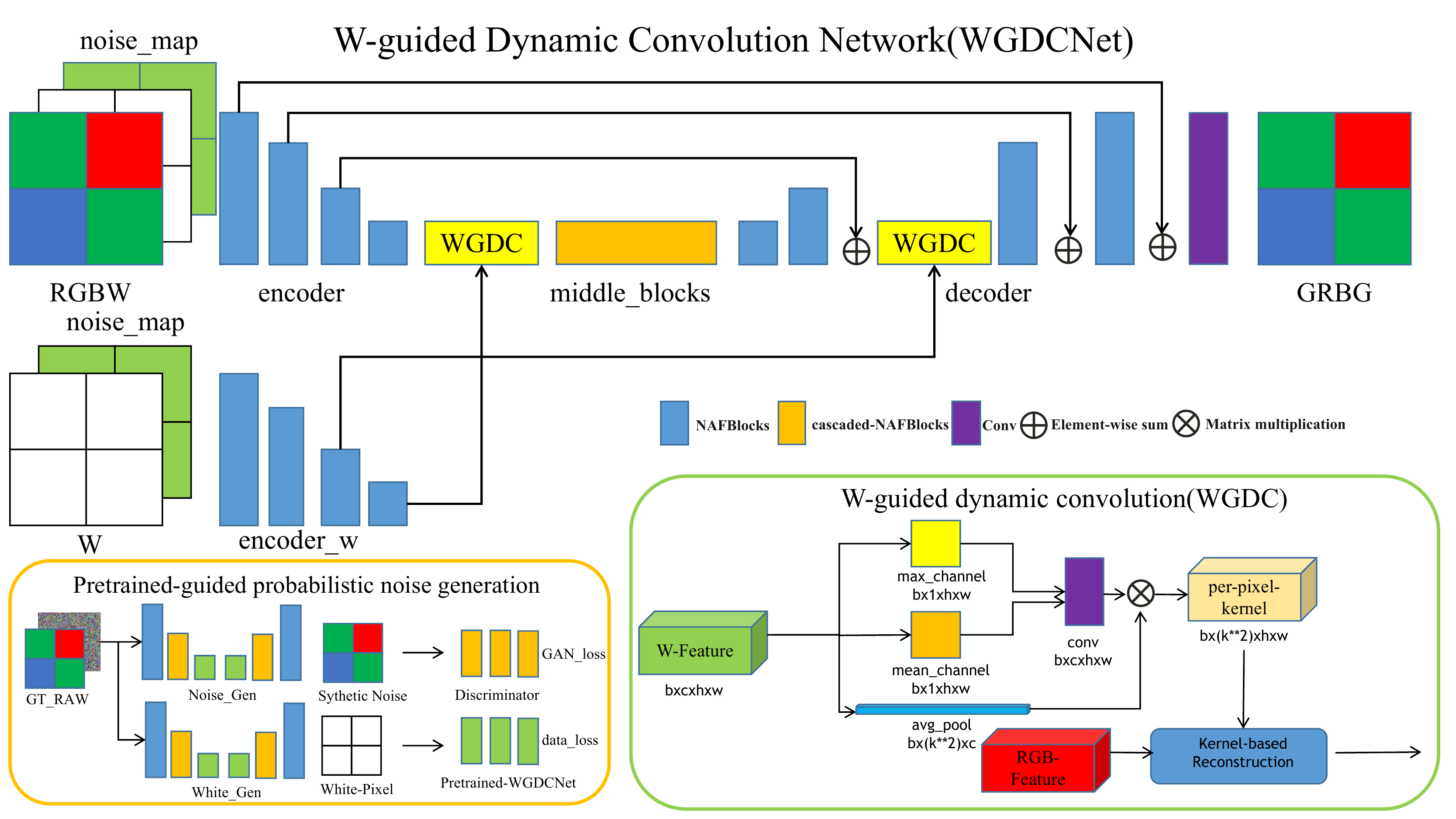}
    \caption{The model architecture of MegNR.}
    \label{fig:WGDCNET}
\end{figure}

This team proposes a novel two-stream pipeline based NAF-blocks~\cite{chen2022simple} for RGBW joint fusion and denoise task, as shown in Fig.~\ref{fig:WGDCNET}. Inspired by high signal-to-noise ratio white pixels, they introduce a new module called W-guided dynamic convolution(WGDC), which aggregates the spatial and channel attributes of the white-pixel feature, guides the dynamic change of network capability according to the white-pixel characteristics. Moreover, the authors design a method for synthesizing RGBW data, which effectively reduces the gap between synthetic data and real data. They use the official simple ISP code to transfer the standard Bayer to the RGB image for loss calculation, which consists of PSNR, SSIM, LPIPS~\cite{lpips}. For training, the authors randomly crop the training images into 128x128-sized patches with the 8-sized batches.  Bayer Preserving Augmentation~\cite{liu2019learning}, Cutmix \cite{yun2019cutmix} are used for data augmentation.  They use cosine decay strategy to decrease the learning rate to $1 \times 10^{-7}$ with the initial learning rate $1 \times 10^{-3}$ and the entire training costs about 5 days and converges after $4 \times 10^{-6}$ iters. In the final inference stage, Test-time Augmentation~\cite{shanmugam2021better} is used to get final result.

\subsection{USTC-Zhalab}

\begin{figure}[!ht]
    \centering
    \includegraphics[width=\linewidth]{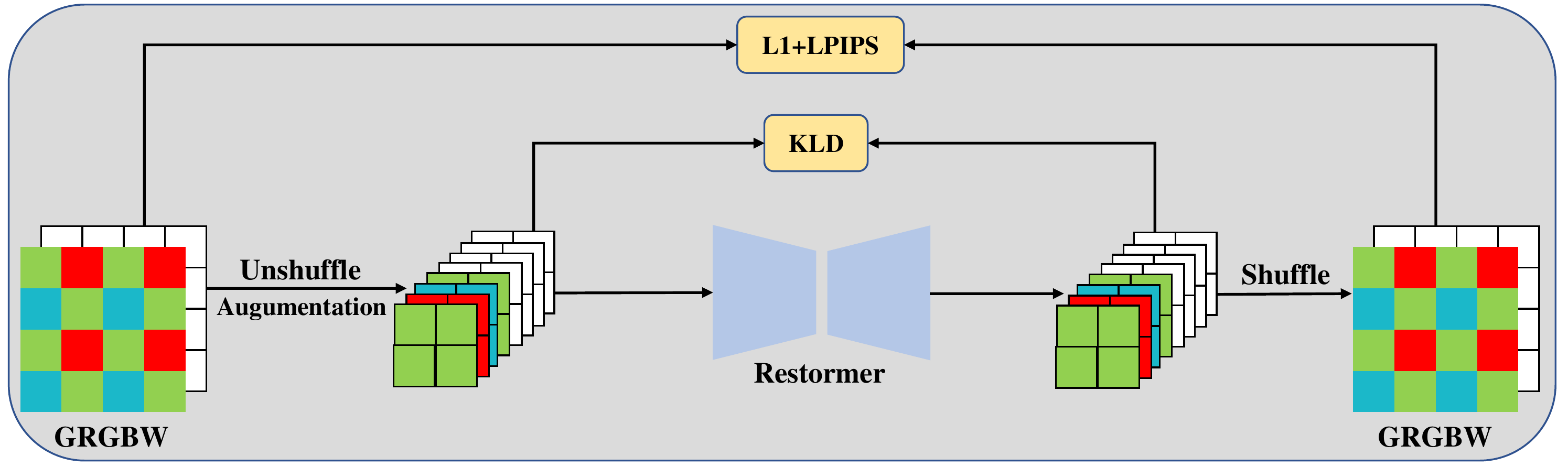}
    \caption{The model architecture of USTC-Zhalab.}
    \label{fig:ustc-zhalab}
\end{figure}

This team proposes an RGBW fusion and denoising method based on the existing image restoration model Restormer \cite{zamir2022restormer}, as shown in Fig.~\ref{fig:ustc-zhalab}. During training, the Pixel-Unshuffle \cite{sun2022hybrid} is firstly applied to RGBW images to split them from 2 channels into 8 channels. Then, the 8-channel RGBW images are fed into the Restormer, obtaining the output of 8 channels. Finally, the Pixel-Shuffle \cite{shi2016real} restores the output of 8 channels to the standard Bayer format. The training loss function consists of L1 loss, KLD loss, and LPIPS loss \cite{lpips}, calculated on the output of 8 channels and GT Bayer. Moreover, they also utilize three data augmentation strategies for training, i.e., Bayer Preserving Augmentation \cite{liu2019learning}, Cutmix \cite{yun2019cutmix}, and Mixup \cite{zhang2017mixup}. The whole network is trained for $3 \times 10^5$ iterations. The learning rate is decayed from $2 \times 10^{-4}$ to $1 \times 10^{-6}$ with a CosineAnnealing schedule. The training batch size and patch size are set to 8 and 224. The Self-ensemble strategy, Test-time Augmentation \cite{shanmugam2021better}, and Test-time Local Converter \cite{chu2021revisiting} are applied during the testing phase. The testing batch size and patch size are set to 1 and 224.

\subsection{VIDAR}
\begin{figure}[!ht]
    \centering
    \includegraphics[width=\linewidth]{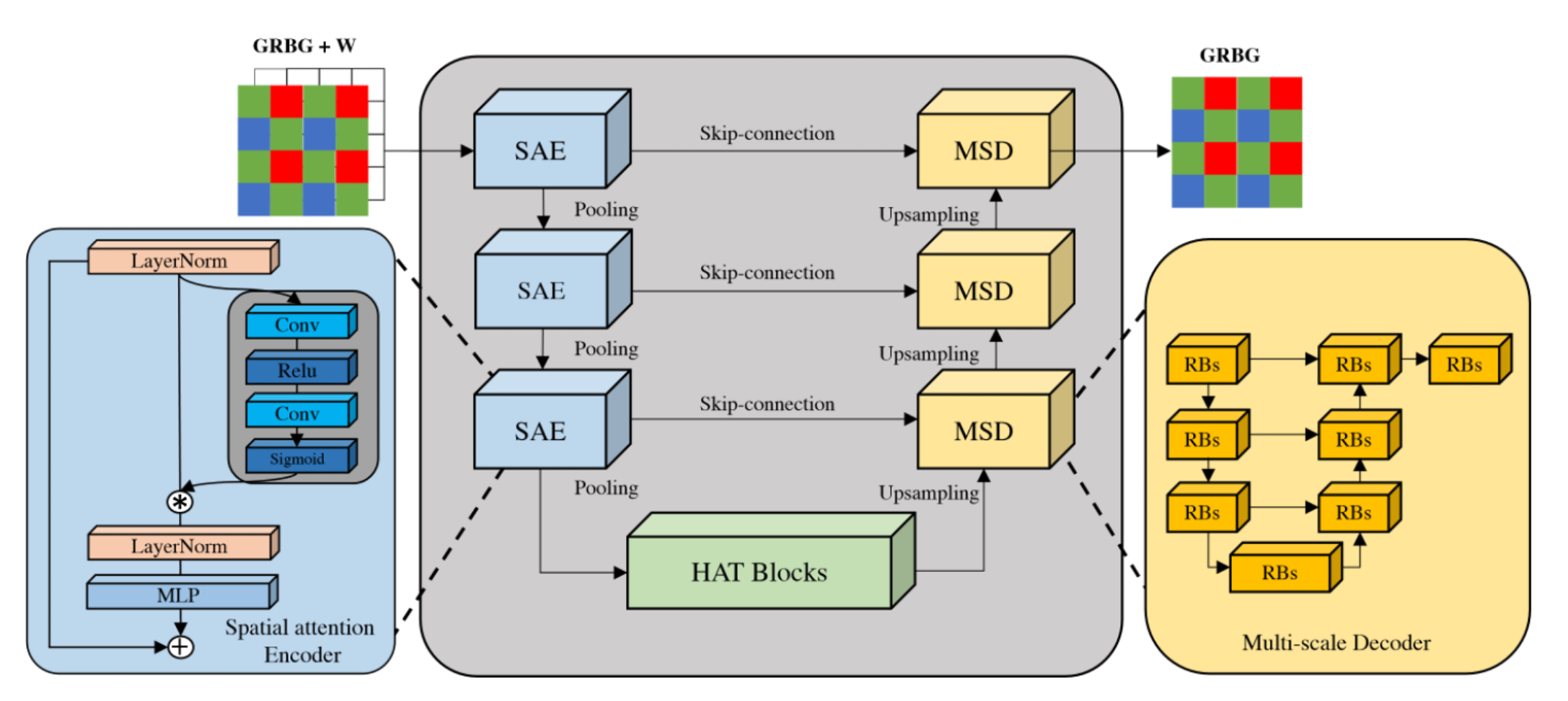}
    \caption{The model architecture of VIDAR.}
    \label{fig:model_vidar}
\end{figure}

This team proposes a multi-scale hybrid attention network for RGBW Fusion and denoising task as shown in Fig.~\ref{fig:model_vidar}. Inspired by Restormer \cite{zamir2022restormer} and HAT \cite{chen2023activating}, the proposed method employs the Spatial Attention Module as the decoder (SAE), which is decoded by the Multi-Scale Decoder (MSD) via skip-connections. The hybrid attention transformer (HAT) is also used in this strategy to consider both global and local information. The combination of these techniques enables efficient processing of high-resolution images as well as the extraction of significant information. The training process includes two stages. In the stage $1$, the network is trained with the patches of size $256\times256$. The batch size is set to $12$ and the optimizer is ADAM by setting of $\beta_{1} = 0.9$ and $\beta_{2} = 0.999$. The learning rate is initialized as $10^{-4}$ and it is decayed by a factor of $0.5$ at $50,000$, $80,000$, and $100,000$ iterations. Thanks to the network in stage $1$ to greatly improve Bayer by using the white-channel information. The stage $2$ finetunes the network from stage $1$. In stage $2$, the network is trained with the $L_{1}$ loss and LPIPS loss on RGB domain. The learning rate is initialized as $5\times10^{-5}$ and decayed by a factor of $0.5$ at $10,000$, $20,000$, and $40,000$ iterations. In the testing phase, the results are from the two stages to achieve the best performance. To be specific, the self-ensemble is used in the period of testing single model to get a better result. The input frame is flipped and regard it as another input. Then an inverse transform is applied to the corresponding output. An average of the transformed output and original output will be the self-ensemble result. The final result is a mixed results of the outputs from different iterations.

\section{Conclusions}
This report reviewed and summarized the methods and results of the RGBW Fusion challenge in the 2nd Mobile Intelligent Photography and Imaging workshop (MIPI 2023) held in conjunction with CVPR 2023. The participants were provided with a high-quality dataset for RGBW fusion and denoising.
The four submissions leverage learning-based methods and achieve promising results. 
We are excited to see so many submissions within such a short period, and we look forward to more research in this area.


\section{Acknowledgements}
We thank Shanghai Artificial Intelligence Laboratory, Sony, and Nanyang Technological University for sponsoring this MIPI 2023 challenge. We thank all the organizers and participants for their great work.


{\small
\bibliographystyle{ieee_fullname}
\bibliography{egbib}
}

\appendix

\section{Teams and Affiliations}
\label{appendix:teams}

\subsection*{RUSH MI}
\noindent
\textbf{Title}:\\ 
Decoupled End-to-End Remosaic and Denoise Mode\\
\textbf{Members}:\\
Hongyuan Yu$^1$ (\href{yuhongyuan@xiaomi.com}{yuhongyuan@xiaomi.com}),\\
Yuqinq Liu$^2$, Weichen Yu$^3$, Lin Ge$^1$,  Xiaolin Zhang$^1$, Qi Jia$^2$, Heng Zhang$^1$, Xuanwu Yin$^1$, Kunlong Zuo$^1$\\
\textbf{Affiliations}:\\
$^1$ Multimedia Department, Xiaomi Inc.,\\
$^2$ Dalian University of Technology,\\ 
$^3$ Institute of Automation, Chinese Academy of Sciences
\\

\subsection*{MegNR}
\noindent
\textbf{Title}:\\ 
WGDCNet: W-guided Dynamic Convolution Network for RGBW Fusion Image\\
\textbf{Members}:\\
Qi Wu$^1$ (\href{wuqi02@megvii.com}{wuqi02@megvii.com}),\\
Wenjie Lin$^1$, Ting Jiang$^1$, Chengzhi Jiang$^1$, Mingyan Han$^1$,  Xinpeng Li$^1$, Jinting Luo$^1$, Lei Yu$^1$, Haoqiang Fan$^1$, Shuaicheng Liu$^{2, 1*}$\\
\textbf{Affiliations}:\\
$^1$ Megvii Technology,\\
$^2$ University of Electronic Science and Technology of China (UESTC)
\\

\subsection*{USTC-Zhalab}
\noindent
\textbf{Title}:\\
Restormer-based RGBW Joint Fusion and Denoise\\
\textbf{Members}:\\
Kunyu Wang (\href{kunyuwang@mail.ustc.edu.cn}{kunyuwang@mail.ustc.edu.cn}),\\
Chengzhi Cao\\
\textbf{Affiliations}:\\
University of Science and Technology of China
\\

\subsection*{VIDAR}
\noindent
\textbf{Title}:\\
Multi-Scale Hybrid Attention Network for RGBW Fusion and Denoising\\
\textbf{Members}:\\
Yuanshen Guan (\href{guanys@mail.ustc.edu.cn}{guanys@mail.ustc.edu.cn}),\\
Jiyuan Xia, Ruikang Xu, Mingde Yao, Zhiwei Xiong \\
\textbf{Affiliations}:\\
University of Science and Technology of China

\end{document}